# Experimentally derived axial stress- strain relations for two-dimensional materials such as monolayer graphene

Ch. Androulidakis[1,2], G. Tsoukleri[2], N. Koutroumanis[2], G. Gkikas[2], P. Pappas[2], J. Parthenios[2], K. Papagelis[1,2] and C. Galiotis[2,3*]

1. Materials Science Department, University of Patras, Greece
2. Institute of Chemical Engineering and High Temperature Chemical Processes, Foundation of Research and Technology-Hellas (FORTH/ICE-HT), Patras, Greece
3. Department of Chemical Engineering, University of Patras, Greece

*Corresponding author: c.galiotis@iceht.forth.gr

**Abstract**

A methodology is presented here for deriving true experimental axial stress-strain curves in both tension and compression for monolayer graphene through the shift of the 2D Raman peak ($\Delta\omega$) that is present in all graphitic materials. The principle behind this approach is the observation that the shift of the 2D wavenumber as a function of strain for different types of PAN-based fibres is a linear function of their Young's moduli and, hence, the corresponding value of $\Delta\omega$ over axial stress is, in fact, a constant. By moving across the length scales we show that this value is also valid at the nanoscale as it corresponds to the in-plane breathing mode of graphene that is present in both PAN-based fibres and monolayer graphene. Hence, the $\Delta\omega$ values can be easily converted to values of $\sigma$ in the linear elastic region without the aid of modelling or the need to resort to cumbersome experimental procedures for obtaining the axial force transmitted to the material and the cross-sectional area of the two-dimensional membrane.

**1. Introduction**

Graphene consists of a two-dimensional (2-D) sheet of covalently bonded carbon and forms the basis of both 1-D carbon nanotubes, 3-D graphite but also of important commercial products, such as, polycrystalline carbon (graphite) fibres. As a single defect-free molecule, graphene is predicted to have an intrinsic tensile strength higher than any other known material [1] and tensile stiffness similar to values measured for graphite. However, to date relatively little experimental work has been published regarding the behaviour of atomically thin membranes such as the monolayer graphene (1 LG) under various types of mechanical loading (tension, compression, bending etc.). Early bending experiments [2] have indeed confirmed the extreme stiffness of graphene of 1 TPa and provided an indication of the breaking strength of graphene of 42 N m$^{-1}$ (or 130 GPa graphene thickness of 0.335 nm). These experiments involved the simple bending of a tiny flake by a nano-indenter and the axial force-displacement response was derived by considering graphene as a clamped circular membrane made by an isotropic material. In that experiment it was assumed-albeit not explicitly stated- that the monolayer graphene behaves like a membrane of practically zero bending stiffness. Such a thin membrane can only sustain tensile loads under a bending configuration and would wrinkle spontaneously under compression. Other investigators have indeed estimated or indirectly measure a small –but not trivial- bending stiffness for 1GL of approximately 1-2 eV estimated from the phonon spectrum [3] and the resonance frequency of graphite [4]. Other authors [5] have considered graphene as a stiff plate and its behaviour has been studied in the light of conventional plate mechanics. However, such an approximation is also laden with problems. To start with, a plate has normally a finite thickness that gives rise to an internal stress/ strain distribution during bending; such an assumption has no physical meaning for a plate of atomic thickness. Furthermore, if we treat the suspended membrane as a thin plate with a Young's modulus $E=1$ TPa, Poisson's ratio $v=0.16$ [6], and thickness $t=0.334$ nm, then from the well-established plate mechanics formula:

$$D = \frac{Et^3}{12(1-v^2)}$$

a bending stiffness $D=20$ eV is obtained [7]. Hence the plate phenomenology seems to be also problematic at least for suspended graphene in air. Today that graphene science has taken off and many applications can find their way into the market, it is of paramount importance to understand fully the mechanical response of 2-D materials such as graphene and to be able to monitor in a reliable manner the axial stress-strain behaviour.

Over the last few years we have published a series of papers [8-11] on the monitoring of the axial deformation of graphene flakes under the imposition of external tension and compression forces. In these experiments we have used beam-type loading systems developed in the early nineties [12] in order to subject the 2-D materials to tension and compression while the molecular deformation was monitored with Raman spectroscopy. This work has confirmed the extreme stiffness of graphene of 1 TPa [9] and have provided an estimate of the compression strain to failure of single flakes embedded in polymer matrices which was found to be independent of its geometrical characteristics. In axial tension, a linear relationship between Raman frequency and strain was established for the monolayer graphene up to strains of about 1.5%. However, due to restrictions of the flexed-beam configuration, these 2-D materials cannot be strained to deformations much higher than 1.5%. In addition, the techniques applied so far involve the recording of the molecular vibrations as a function of strain and provide no direct information on axial force which is required for obtaining true axial stress (force) vs strain (displacement) curves.

In this paper we compare the phonon deformation of graphite/ graphene crystals in carbon fibres with those of monolayer graphene and we construct a universal map of 2D graphene phonon deformation as a function of Young's modulus. The results obtained across the lengths scales allow us to transform the shift of the phonon wavenumber, $\Delta\omega$, per increment of strain to values of $\Delta\omega$, per increment of stress and, thus, to transform the Raman wavenumber vs. strain curves to true axial stress-strain relationships for graphenes embedded into polymer matrices. Since axial force-displacement experiments are very difficult –if not impossible- to perform on 2-D materials, this appears to be a viable method to produce axial stress-strain curves for tensile deformations not exceeding a few per cent of strain. Moreover, in compression for which failure is indeed observed at less than 1% [13], a full stress-strain relationship up to failure (and beyond) can be established.

## 2. Experimental

**Carbon Fibres**

Two types of High Modulus (HM) PAN based Carbon Fibres with E ≈ 540GPa (M55J) and E ≈588GPa (M60J) respectively were tested in this study. Both fibres were 5 μm in diameter and were supplied by Toray Industries in 6K tows. Single fibres were separated from the tow and aligned axially in 25mm gauge length paper frames using a commercial two component epoxy resin. Micro-Raman spectra were measured using two different lines of laser, 514.5 nm (2.41eV) and 488nm (2.54eV) respectively. The laser power was kept below 1.1mW on the fibre in order to avoid local overheating. A 80x objective with numerical aperture 0.75 was used and the spot size of the laser on the fibre was estimated to approximately 2μm$^2$. A triple monochromator was employed as a phonon counting system to collect the back scattering data. All the Raman frequency values were derived by fitting Lorentzian routines to the charge coupled device (CCD) raw data. Individual CFs on the paper frames were transferred to the jaws of a small staining device, with their axes aligned parallel to the stretching direction to ±5°. The fibre extension measured to ±1μm and the strain was increased in steps 0.04-0.12% up to failure 0.5-0.9%. The spectra were taken at the middle of the fibre and three measurements were averaged at each step. **Figure 1** shows the Raman spectra obtained in the region of 2550 to 2950 cm$^{-1}$ (2D peak) for the examined carbon fibres. The 2D band consists of two-phonon combination and therefore requires the existence of a certain degree of order in order to be present. The slopes were calculated using a least-squares-fit to the data and the results are shown in **figure S3**.

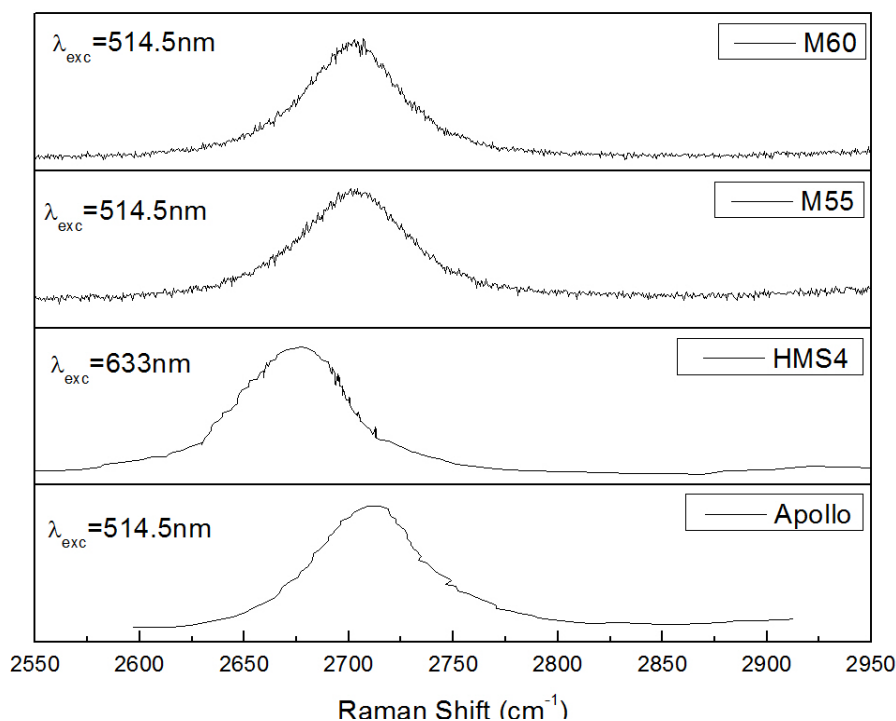

**Fig. 1 Representative Raman spectra of the 2D peak for the examined CF. The M55 and M60 were supplied by Toray Industries whereas the HMS4 and Apollo fibre were supplied in the past by Hercules Inc. (US) and Courtaulds (UK), respectively.**

**Graphene**

Graphene samples were prepared by mechanical exfoliation of HOPG. The samples were deposited on PMMA bars and the number of layers was identified using Raman spectroscopy. On the top of the samples another PMMA layer of thickness ~100 nm was spin coated in order to create a sandwich sample for efficient stress transfer from the polymer to the graphene. By bending the PMMA bars using a four-point-bending jig, stress induced to the graphene flakes. The strain was applied incrementally and Raman measurements were taken *in situ* at every loading step. The laser excitations that used were the 514 nm and 785 nm. Analytical details for the experimental procedure can be found in Refs [8, 10, 13].

## 3. Results and Discussion

It is well established, specifically for graphitic materials, that the position of the 2D and G Raman peaks, shift under applied mechanical strain [9, 14-16]. In a previous paper [9] we looked at the behaviour of the doubly degenerate $E_{2g}$ peak and established a universal relationship for the peak shift that is valid for graphene and a whole range of PAN-based carbon fibres that exhibit an onion skin morphology (like giant nanotubes). The D peak, which is normally present in carbon fibres, is due to the breathing modes of $sp^2$ rings and requires a defect for its activation [17]. It comes from the transverse optical phonon branch (TO) around the K point of the Brillouin zone [18, 19] is active by double resonance (DR) [20] and is strongly dispersive with excitation energy due to a Kohn Anomaly at K [12]. It is also considered to be a similar breathing mode to the TO $A_{1g}$ phonon at K [21]. For a pure $A_{1g}$ symmetry and relatively small strains (<2%), $\Delta\omega_D$, the uniaxial shift in graphene, is related solely to the hydrostatic component of the strain [22] such as:

$$\Delta\omega_D = -\omega_D \gamma_D \left(\varepsilon_{ll} + \varepsilon_{tt}\right) \quad \textbf{or}$$
$$\Delta\omega_D = -\omega_D \gamma_D \varepsilon \left(1 - \nu_g\right) \quad (1)$$

where $\varepsilon_{ll}$ and $\varepsilon_{rr}$ are the longitudinal and transverse strains ($\varepsilon_{tt} = -\nu_g \varepsilon_{ll}$), $\omega_D$ is the *D* wavenumber at rest, $\gamma_D$ is the Gruneisen parameter for that mode and $\nu_g$ is the axial-transverse Poisson's ratio of graphite that has been reported as ranging from 0.13 to 0.20 [6, 23-25]. For experiments conducted on graphene flakes the applied strain, $\varepsilon$, is identical to $\varepsilon_{ll}$.

The 2D peak is a two phonon overtone of the D peak mentioned above. It is a single peak in monolayer graphene, whereas it splits in multiple bands in bilayer and trilayer graphene, reflecting the evolution of the band structure [26]. Therefore, for a suspended graphene flake in air equation 1 can be written as:

$$(\Delta \omega_{2D})_{air} = -\omega_{2D} \gamma_{2D} \varepsilon (1 - v_g) \qquad (2)$$

Where $\Delta\omega_{2D}$ is the shift of the 2D peak that results from the hydrostatic component of the strain and $\gamma_{2D}$ is the Gruneisen parameter. It should be stressed that 2D and D are laser dispersive modes with slopes of about 100 cm$^{-1}$/eV and 50 cm$^{-1}$/eV, respectively [21]. As a result, any changes in the laser excitation, or in the Poisson's ratio (in case graphene is embedded in or attached to a matrix) is bound to alter the value of $\omega_{2D}$, and hence the measured shift per strain and this may explain, in part, observed discrepancies in the literature regarding the strain sensitivity of the 2D peak [8, 9, 15, 22].

For polycrystalline graphitic materials such as carbon fibres the situation is more complex. The uniaxial shift in carbon fibres is related to the hydrostatic component of the strain [22] which in this case is given by:

$$(\Delta \omega_{2D})_{air} = -\omega_{2D} \gamma_{2D} (\varepsilon_{ll} + \varepsilon_{\theta\theta} + \varepsilon_{rr}) \quad \textbf{or}$$
$$(\Delta \omega_{2D})_{air} = -\omega_{2D} \gamma_{2D} \varepsilon_{ll} (1 - v_{\theta\theta} - v_{rr}) \qquad (3)$$

Where $\varepsilon_{ll}$, $\varepsilon_{\theta\theta}$, $\varepsilon_{rr}$ are the longitudinal, hoop and radial strains and $v_{\theta\theta}$ and $v_{rr}$ are the axial-hoop and axial-transverse Poisson's ratios. It is worth adding here that in this case the $\varepsilon_{ll}$ is not identical to the applied axial strain $\varepsilon$ due to crystallite slippage and rotation that do not contribute to $\varepsilon_{ll}$. By differentiating (3) with respect to applied strain $\varepsilon$ we obtain:

$$\frac{d(\Delta \omega_{2D})_{air}}{d\varepsilon} = -\omega_{2D} \gamma_{2D} (1 - v_{\theta\theta} - v_{rr}) \frac{d\varepsilon_{ll}}{d\varepsilon} \qquad (4)$$

Past work by our group and others [27-29], has shown that polycrystalline fibres such as carbon and aramid are equal-stress materials (ie springs-in-series) and therefore for a given applied (axial) stress the shift of the Raman peaks in the first order region of the spectrum is the same regardless of modulus. This hypothesis has been verified by independent strain controlled and stress controlled experiments for PAN based carbon fibres [29] and more recently [9] the stress sensitivity of the G peak in air for carbon fibres has been linked to the corresponding stress sensitivity in monolayer graphene. Indeed, when the Raman wavenumber shift is measured over the applied strain, non-linear effects such as crystallite slipping or rotation affect the strain in the fibre but do not contribute to bond extension (or contraction) hence the Raman wavenumber is not affected. On the other hand, when the Raman wavenumber is scaled to stress a constant value is obtained since non-linear mechanisms as above do not affect the Raman measurements. Thus by eliminating $\Delta v$ from $\Delta v = f(\varepsilon)$ and $\Delta v = f(\sigma)$ equations a true $\sigma = f'(\varepsilon)$ relationship can be derived in both tension and compression (see Supporting Information). In monolayer graphene for relatively small deformations (~1.5%) the tensile stress-strain relationship is linear and therefore, as it will argue below, the conversion of the 2D wavenumber shift per strain to the equivalent value per stress is quite straightforward. A detailed explanation of past work in the area is presented in the Supporting Information.

For small strains and linear behaviour ($\varepsilon = \sigma/E$) equation (4) can be written as:

$$\left(\frac{\Delta \omega_{2D}}{\varepsilon}\right)_{air} = -\omega_{2D} \gamma_{2D} \left(1 - v_{\theta\theta} - v_{rr}\right) \frac{E_f}{E_g} \quad \text{or}$$

$$\left[\frac{1}{\omega_{2D}}\right]\left(\frac{\Delta \omega_{2D}}{\varepsilon}\right)_{air} = -\left[\frac{\gamma_{2D}\left(1 - v_{\theta\theta} - v_{rr}\right)}{E_g}\right] E_f \quad \text{or} \quad (5)$$

$$\left(\frac{\Delta \omega_{2D}}{\omega_{2D}}\right)_{air} = -\left[\frac{\gamma_{2D}\left(1 - v_{\theta\theta} - v_{rr}\right)}{E_g}\right] \sigma$$

Where $E_f$ and $E_g$ are the fibre and graphene Young's moduli, respectively. The above equations provide for the first time an analytical expression for the experimentally verified linear relationship between the 2D wavenumber shift and the axial stress for all PAN based fibres [28]. It states that for carbon fibres loaded in air, the normalized wavenumber shift relates linearly to stress. To verify this we plot in **figure 2** the wavenumber shift ($\Delta\omega_{2D}$) as a function of strain, for a whole range of highly crystalline PAN-based carbon fibres of various Young's moduli, $E_f$, loaded in air.

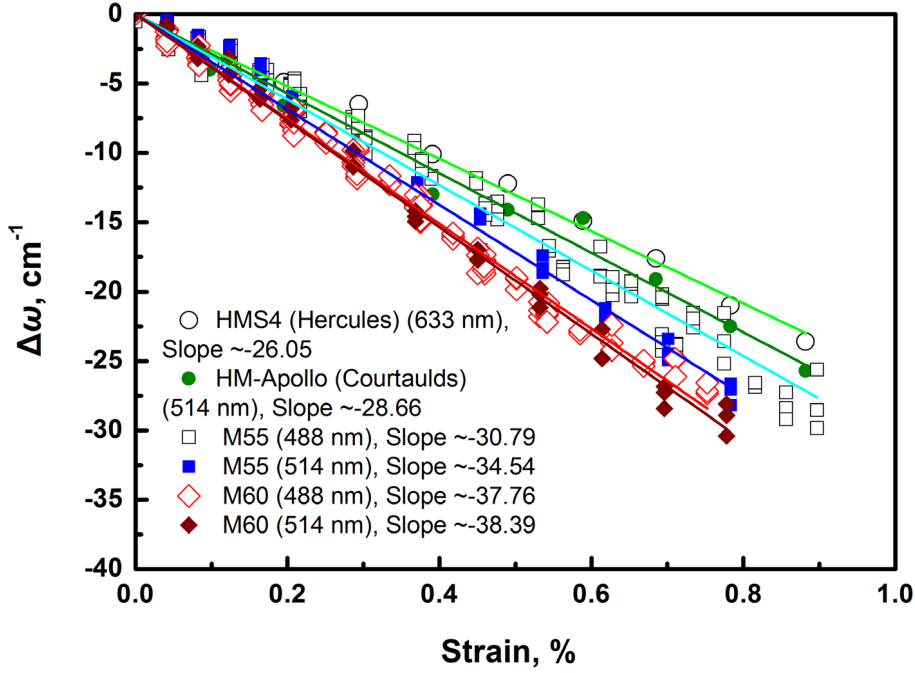

*Fig. 2.* **Wavenumber shift vs. Young's modulus for a whole range of PAN-based carbon fibres in air.**

Furthermore in **figure 3** we plot the normalized wavenumber shift per strain as a function of $E_f$ (over $E_g$). As is evident, all carbon fibre experimental points lie on a least-squares-fitted straight line of slope of -2.40 and by assuming $\gamma_{2D}$=3.55 a value of 0.68 is obtained for the Poisson's ratio expression of equation (5) which amounts to $v_{rr}+v_{\theta\theta}$=0.32. This is a reasonable value for PAN-based carbon fibres with an onion-skin structure [30]-[31]. It is worth adding here that the relationship between $\Delta\omega_{2D}/\varepsilon$ and $\omega_{2D}$ (the value of the 2D wavenumber at rest) is also confirmed by the experimental data of Fig.2 since the slopes of the lines for a specific fibre increases as the $\omega_{2D}$ increases. The latter is inversely proportional to the exciting wavelength due to the observed Kohn anomaly at K [32], hence, one needs a low excitation wavelength in order to maximize the strain/stress sensitivity.

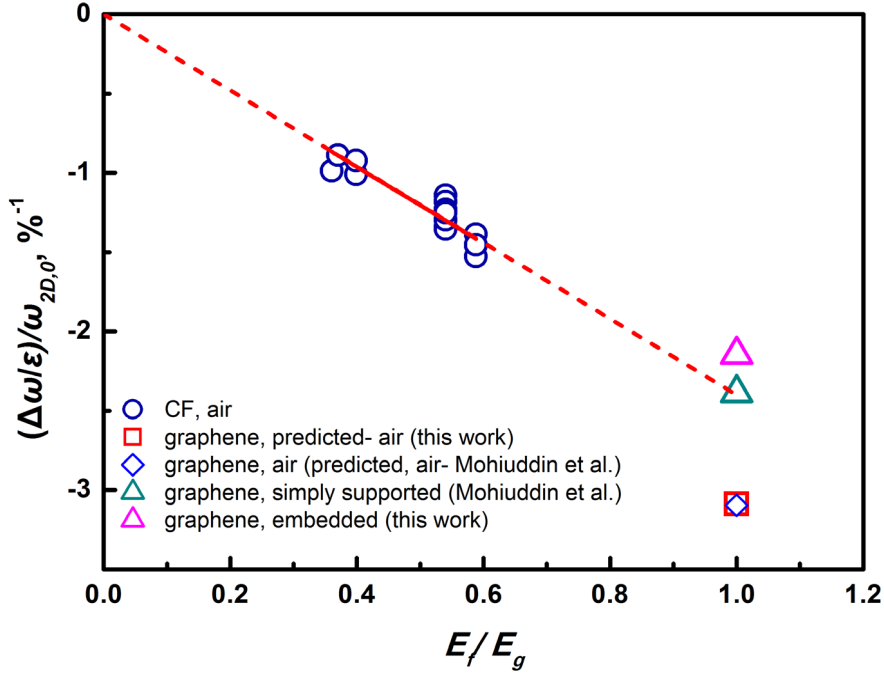

*Fig. 3.* **Wavenumber shift per strain normalised by the 2D wavenumber (at zero strain) vs. axial Young's modulus for a whole range of PAN-based carbon fibres. The solid line with a slope of -2.41 is least-squares-fitted to the carbon fibre experimental points. The dotted line is extrapolation to the graphene region for which $E_f \equiv E_g = 1$.**

A number of experiments have been also performed on graphenes fully embedded into polymer systems such as SU8/PMMA systems of various thicknesses. By pursuing a similar analysis to that presented above for carbon fibres (eq. (5)), we obtain (equation (2)) for graphene either free-standing in air or embedded in a polymer matrix, the following expressions:

$$\left[\frac{1}{\omega_{2D}}\right]\left(\frac{\Delta\omega_{2D}}{\varepsilon}\right)_{polymer} = -\left[\frac{\gamma_{2D}(1-\nu_{SU8/PMMA})}{E_g}\right]E_g \quad \text{and} \quad \left[\frac{1}{\omega_{2D}}\right]\left(\frac{\Delta\omega_{2D}}{\varepsilon}\right)_{air} = -\left[\frac{\gamma_{2D}(1-\nu_g)}{E_g}\right]E_g$$

or

$$\left(\frac{\Delta\omega_{2D}}{\omega_{2D}}\right)_{polymer} = -\left[\frac{\gamma_{2D}(1-\nu_{SU8/PMMA})}{E_g}\right]\sigma \quad \text{and} \quad \left(\frac{\Delta\omega_{2D}}{\omega_{2D}}\right)_{air} = -\left[\frac{\gamma_{2D}(1-\nu_g)}{E_g}\right]\sigma$$

(6)

Where $\nu_{SU8/PMMA}$ is the Poisson's ratio of the polymer system. By plucking the value of -64.0 cm$^{-1}$/% measured earlier on a simply supported SU8/PMMA we obtain as before $\nu_{SU8/PMMA}$=0.33. The results reported here (**figure 4a**) have yielded average values of -54.0cm$^{-1}$/% and -57.5cm$^{-1}$/%in tension under 785 nm and 514 nm excitations, respectively. This corresponds to values of $\nu_{SU8/PMMA}$=0.39-0.40 which are quite reasonable for the SU8/ PMMA system [33]. If we restrict ourselves to the linear region of the stress-strain curve $\varepsilon=\sigma/E_g$ therefore the wavenumber shift normalised by the excitation line is proportional to applied stress similarly to the relationship obtained for carbon fibres (equation (5)) regardless of Young's modulus. For $\omega_{2D}$=2595 cm$^{-1}$ (785 nm excitation to avoid matrix fluorescence) we get wavenumber per stress rates of -5.5 cm$^{-1}$ GPa$^{-1}$. For 514 nm excitation ($\omega_{2D}$=2680 cm$^{-1}$), the corresponding value is -5.7 cm$^{-1}$ GPa$^{-1}$. It is worth noting that these values compare reasonably well with the value of 6.4 cm$^{-1}$ GPa$^{-1}$ reported by Mohiudin *et al.* [22] obtained for 514 nm excitation but for a simply-supported specimen (reduced Poisson's effect).

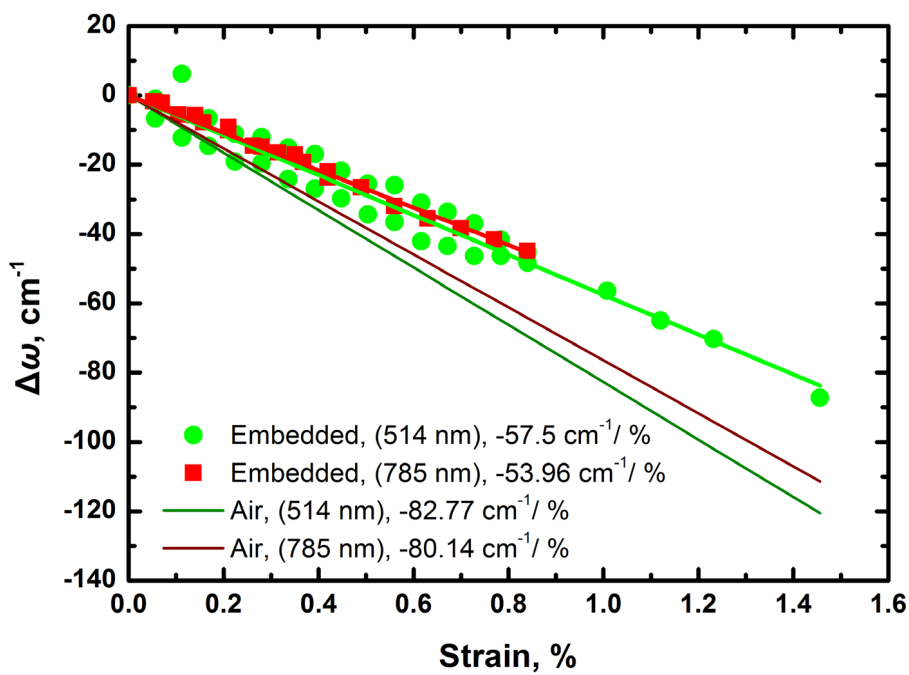

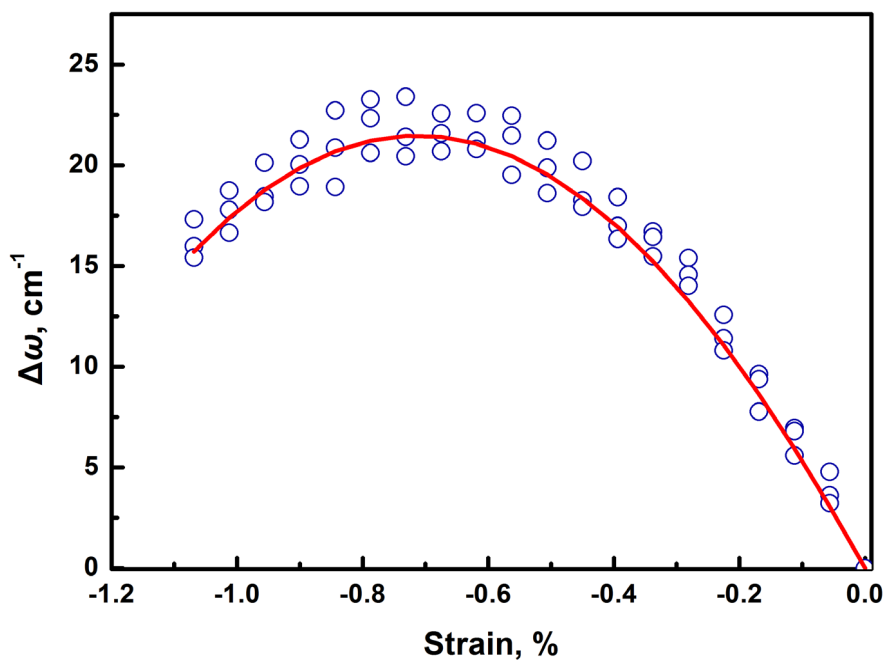

*Fig. 4.* **Wavenumber shift vs. strain for monolayer graphene for three independent measurements on large (>10 μm length) flakes. *(a)* Axial tensile loading: The solid lines is least-squares-fitted to the data points (slope=56.90 cm$^{-1}$/%). The expected lines for deformation in air for an excitation**

**wavelength of 785 nm are also presented (b) Axial compression loading: The solid line is a fourth degree polynomial fitted to the experimental data and the open circles are Raman measurements for a flake with length of ~30 μm.**

Representative results for graphene in compression are presented in **figure 4b**. In this case, the curve of the wavenumber shift vs. strain is non-linear and exhibits a plateau at approximately -0.6%. As presented elsewhere [10], the 2-D monolayer in air has effectively no resistance to compression loading; however, embedded into the SU8/PMMA matrix the monolayer is restricted from buckling till interface yielding or failure in the lateral direction allows the formation of a sinusoidal wave of estimated wavelength of about 1-2 nm and a height of 0.7 nm. This behaviour is very different than the response of carbon fibres of microscopic dimensions to axial compression for which prior to Euler (elastic) buckling the material fails by shear or bulging [12]. In effect, 2-D materials well supported by surrounding matrices provide a more effective reinforcement in spite of their atomic dimensions. In a future publication, the effect of increasing the material thickness through the addition of graphene layers upon the compression behaviour of multi-layer graphenes will be examined.

As mentioned earlier, the wavenumber shift per stress for both carbon fibres and graphene is a constant that is related to the Gruneisen parameter and the modulus of graphene which are common in both. The only notable difference between carbon fibres (macroscale) and graphene (nanoscale) is the Poisson's expression which reflects differences in their morphologies and the environment in which measurements are made (air or polymer). In **figure 3** we have added the values for monolayer graphene ($E_f \equiv E_g$) as predicted for measurements in air and those obtained experimentally from the fully embedded specimens (equation (6)). As expected, the data points for measurements conducted in air are markedly different. This is to be expected due to the differences in the morphology of graphene monolayer and the graphene (graphite) units in the polycrystalline PAN-based carbon fibres. However, it is interesting to note that the data points corresponding to embedded graphene are broadly lying close to the carbon fibre line. This is, indeed, not surprising since, in this case, the Poisson's expression (eq. (6)) yields a value of about 0.6 which compares fortuitously well with the corresponding value of 0.68 for carbon fibres (measurements in air).

We can now turn attention to the use of the stress sensitivity of the 2D phonon to convert spectroscopic data into values of stress (in GPa) for the fully embedded graphene. In **figure 5**, the 2D phonon frequency as a function of strain is converted to an axial stress-strain curve in both tension and compression by employing the stress sensitivity of the 2D wavenumber as mentioned above.

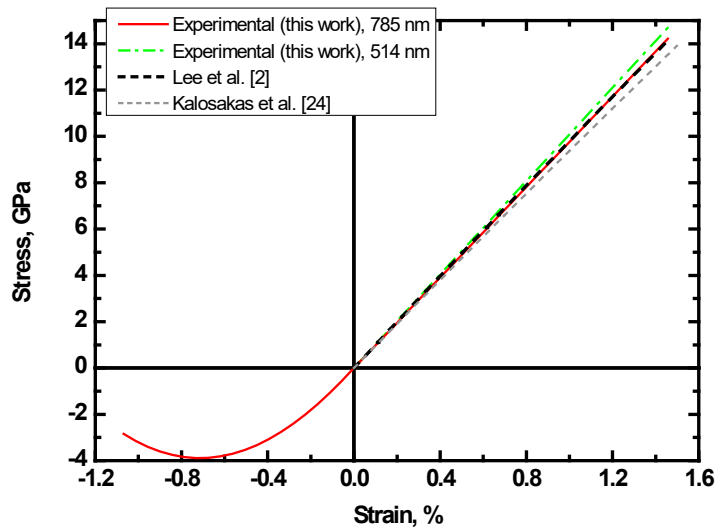

*Fig. 5.* **Experimentally derived axial stress-strain behaviour of monolayer graphene in both tension and compression. The slope in tension corresponds to a tensile modulus of ~0.97TPaand of ~1.01 TPa for 785 nm and 514 nm excitation respectively. No failure is observed in tension up to 1.5% strain corresponding to a stress of 15 GPa. The modulus dependence on strain in compression is shown in Fig. 5. Failure in the form of graphene buckling (wrinkling) is observed at a value of -4 GPa.**

As seen, the data in tension are quite linear up to a value of strain of 1.5% that is the limit of our experimental apparatus. No material failure is observed up to that strain level that corresponds to a stress of ~15 GPa. Indeed this value is already 3 times higher than the tensile strength of the strongest commercially available carbon fibres. In compression a 2D material such as a monolayer graphene is expected to fail in air at strains, as small as 1 nanostrain due to its almost zero thickness (Euler elastic buckling) [10]. However, when embedded in polymer matrices both sides of the monoatomic membrane are prevented from out-of-plane deformation by the presence of the polymer. When the lateral van den Waals bonds eventually yield or fail at a critical lateral strain then the whole or part of the monolayer wrinkles and no further axial stress can be sustained. As seen in **figure 5** this corresponds to a strain of -0.6% and a maximum axial stress of ~-4 GPa. Again this value is comparable to the compressive strength of carbon fibres which fail by shear or bulging in spite of possessing a cross-sectional area 4 orders of magnitude larger (typical value for a CF $10^7$ nm$^2$ as compared to a $10^3$ nm$^2$ for 1LG). This confirms the advantage -per unit of mass- of embedded 2-D materials under compression that are not amenable to shear failure, as compared to the behaviour of commercial fibres such as carbon or aramid. The possibility of producing new composite architectures with exceptional compression properties using sheets of monolayers is currently under investigation.

It is interesting to compare now the curves obtained from spectroscopic data with that derived from the nano-indentation experiment of a suspended graphene sheet employing an empirical non-linear equation of the form $\sigma = E\varepsilon + D\varepsilon^2$, where $E$ is the Young's modulus and $D$ is the third order elastic modulus, respectively [2]. On the same graph, we also plot results of modelling the in-plane motion of graphene sheets by employing bond stretching and angle bending force fields. As presented elsewhere [25], the obtained force fields derived using first principles calculations, providing efficient means of calculations in molecular mechanics simulations. As can be seen up to 1.5 % strain, the experimental data on embedded graphene compare well with those obtained from bending experiments on suspended sheets and also those obtained by modelling using empirical force fields that are input into molecular dynamics simulations. Deviations between the various approaches are expected at higher strains for which the assumptions of the bending experiment are expected to break down. Equally the axial experiments presented here depend on the mechanical integrity of the surrounding matrix at high strains (up to ~30%) which is not normally attainable for glassy polymers. However, fully embedded large sheets into elastomeric matrices have the potential to reach the required strains in order to confirm or refute the predicted failure strain of graphene at 30% [25] and at a tensile strength of ~100 GPa.

Finally, in **figure 6** the Young's modulus as a function of applied axial strain is plotted in tension and compression for both 514 nm and 785 nm laser excitations. As seen, for the tensile measurements up to 1.5% strain the Young's modulus is constant at about ~1 TPa. However, in compression the behaviour is not linear up to first failure at -0.6% strain that corresponds to graphene wrinkling as examined elsewhere [13]. The non-linear behaviour possibly reflects the slight eccentricity in applying an axial load to a monolayer sheet and/or it is a consequence of the non-linear response of the lateral "springs" that prevent graphene from buckling collapse during compression loading. At any rate, the confirmed validity of the stress dependence of the 2D wavenumber regardless of the type of loading (tension or compression) for all graphitic materials allow us to interpret the non-linearity in compression of monolayer graphene (that leads to a linear decrease of Young's modulus, **figure 6**) as a geometric and/or interface problem and not a material characteristic. Work is under way to affirm this assertion by subjecting graphene to cyclic loading and to perform experiments in different matrices.

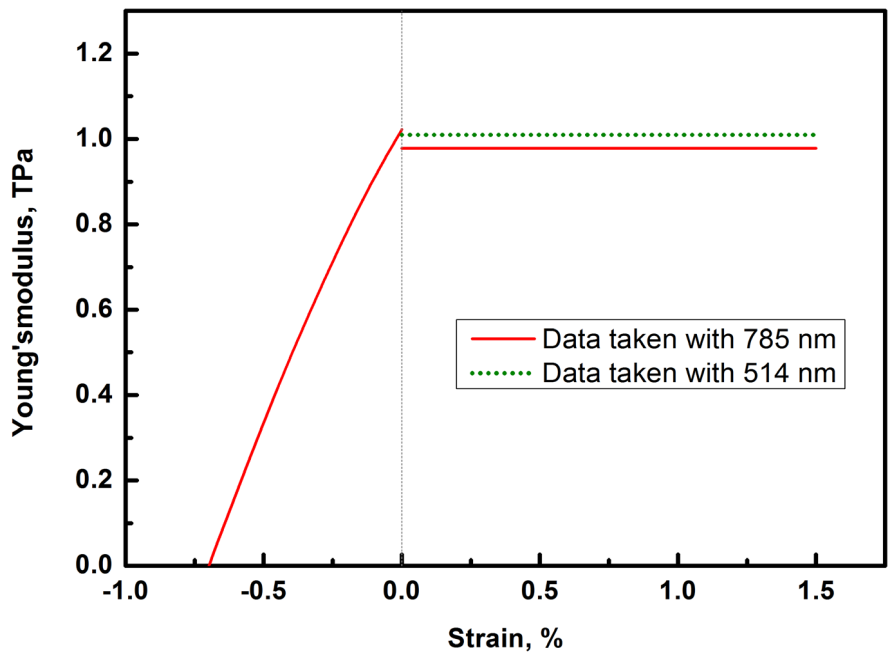

*Fig. 6.* **Strain dependence of graphene Young's modulus as derived from the spectroscopic presented here (for 785 nm and 514 nm excitations). The observed**

**mismatch at zero strain is due to the slightly different Raman strain sensitivities measured in tension and compression.**

4. Conclusions

The experimentally verified observation that the Raman wavenumber shift of a whole range of carbon fibres is linearly proportional to the axial stress is employed here to monolayer graphene through the use of the 2D peak Gruneisen parameter which are common to both (CF and graphene). Care was exercised to compare results taken with the same laser line as the excitation frequency affects the value of the 2D wavenumber at rest of all graphitic materials. From the slopes of the normalised wavenumber per applied strain and the Gruneisen parameter we estimated the value of the Poisson's ratio for both carbon fibre and graphene embedded into a polymer (SU8/ PMMA) but also in air. The values obtained compared well with the listed values of Poisson's ratios for both polymer but also carbon fibres and graphene in air. Finally, the estimated stress-strain sensitivity of the 2D peak was employed to convert the spectroscopic data to true axial stress-strain curves in both tension (up to 1.5%) and compression (up to failure). This methodology can be extended to any two dimensional material and represents the only available method to date to derive axial stress-strain curves for these materials for which mechanical testing at the nanoscale are difficult to perform by conventional means.

**Acknowledgements**

The authors acknowledge financial support from the Thales Project "GRAPHENECOMP", co-financed by the European Union (ESF) and the Greek Ministry of Education, also from the ERC Advanced Grant no321124 "Tailoring Graphene to Withstand Large Deformations" and the FP7 "NEWSPEC" programme no 604168 "New cost-effective and sustainable polyethylene based carbon fibres for volume market applications". The Toray Industries (Japan) are thanked for supplying the high-modulus PAN-based carbon fibres. Finally, Dr. D. J. Sfyris is thanked for assistance with some of the calculations presented here.

Open Access funded by European Research Council.  Under a Creative Commons license.

Ch. Androulidakis, G. Tsoukleri, N. Koutroumanis, G. Gkikas, P. Pappas, J. Parthenios, K. Papagelis, C. Galiotis, Experimentally derived axial stress/strain relations for two-dimensional materials such as monolayer graphene, Carbon 81 (2014) 322-328, doi: 10.1016/j.carbon.2014.09.064